\documentclass[11pt]{article}
\usepackage{amsmath, amssymb}
\usepackage{slashed}
\usepackage{verbatim}
\usepackage{latexsym}
\usepackage{euscript}
\usepackage{amsfonts}
\usepackage{verbatim}
\usepackage[]{array}
\usepackage[]{mathrsfs}
\usepackage{color}
\usepackage{graphicx}
\usepackage{amsmath}
\usepackage{amssymb}
\usepackage{amsxtra}
\usepackage{color}
\def\be{\begin{eqnarray}}
\def\ee{\end{eqnarray}}
\def\0{\nonumber}

\def\tr{{\rm tr}}

\def\det{\rm det}
\usepackage{slashed}
\usepackage{verbatim}
\usepackage{latexsym}\usepackage{color}
\usepackage{euscript}
\newcommand\EW{\EuScript{W}}
\newcommand\ER{\EuScript{R}}

\newcommand\EC{\EuScript{C}}\normalfont\large
\begin{document}
\begin{flushright}
{\it (improved and updated version,  Jan. 2020)}
\end{flushright}

\vskip 2cm

\begin{center}

{\LARGE On the trace anomaly for Weyl fermions }

\vskip 1cm

{\large  L.~Bonora$^{a}$ and R.~Soldati$^{b}$
\\\textit{${}^{a}$ International School for Advanced Studies (SISSA),\\Via
Bonomea 265, 34136 Trieste, Italy \\ and INFN, Sezione di
Trieste \\ }
\textit{${}^{b}$ Dipartimento di Fisica e Astronomia, Via Irnerio 46, 40126 
Bologna, Italy\\  and
INFN, Sezione di Bologna\\}}

\vskip 1cm

\end{center}

\vskip 1cm {\bf Abstract.}
This note is a comment on recent papers that have raised a controversy about the 
existence of the odd-parity trace anomaly in a four-dimensional theory of Weyl 
fermions. Without going into too technical details we explain why their methods 
 cannot detect it.

\begin{center}

{\tt Email: bonora@sissa.it, roberto.soldati@bo.infn.it}

\end{center}

\vskip 3cm

\subsection*{Introduction}

The growing attention on conformal field theories in recent years has brought 
about a renewed interest in the so-called trace anomalies. Several researchers 
have been induced to reconsider and sometimes refine calculations and results 
obtained thirty-forty years ago. One of these is the trace anomaly for chiral 
fermions in 4d. In a series of papers \cite{BGL,BDL,BCDDGS,BCDGPS} - see also 
\cite{Nakayama} -  various methods and different 
regularizations were employed to reach the conclusion that  the trace anomaly for a Weyl 
fermion contains, beside an even parity part, which is a combination of the Weyl 
tensor square density
\be
\EW^2=\ER_{nmkl} \ER^{nmkl}-2 \ER_{nm}\ER^{nm} +\frac 13 \ER^2\label{weyl}
\ee
and the Gauss-Bonnet (or Euler) density,
\be
E=\ER_{nmkl} \ER^{nmkl}-4 \ER_{nm}\ER^{nm} + \ER^2,\label{gausbonnet}
\ee
also an odd parity part, the Pontryagin density,
\be
P=\frac 
12\left(\epsilon^{nmlk}\ER_{nmpq}\ER_{lk}{}^{pq}\right)\label{pontryagin}
\ee 
More precisely,  the trace of the energy-momentum tensor at one loop for a Weyl 
fermion is
\be
T_\mu{}^\mu= \frac{1}{180\times16\pi^{2}}\left(\frac {11}4 \,  E- \frac 92\, 
\EW^2 \pm i \frac {15}4\,  P\right)\label{emtracechiral}
\ee  
The $\pm$ sign on the RHS depends on the chirality, opposite chiralities leading 
to opposite signs 
(anomalies with opposite signs for opposite chiralities will be hereafter called 
{\it split anomalies}).
The parity even and dispersive-like coefficients are well-known at one-loop and 
they are not in question here. The  parity odd term (\ref{pontryagin}), although 
already implicitly contained in \cite{CD2}, was instead somehow unexpected,  
mainly because it is unmistakably  imaginary, i.e., of an absorbitive nature. {This suggests its origin can be a cut discontinuity in the effective action.}
As pointed out in \cite{BGL} it may imply a loss of unitarity and, therefore, 
an obstruction or, anyhow, a novel understanding in the coupling of Weyl fermions to gravity. 

These results have been subsequently challenged by other authors, \cite{Bast1, 
Bast2,FZ}, who, using different approaches, found
the non-occurrence of a parity-odd trace anomaly for a Weyl fermion 
coupled to an external metric.

We believe that in order to assess such results it is, first of all, necessary to be acquainted with the different methods used in the different derivations. Our purpose here is to explain these differences and, in particular, how they become crucial when Weyl fermions are involved. 
Our Pole Star in this comment is that Weyl fermions have a definite chirality and this feature has to be preserved (and never forgotten) throughout the calculation. We also wish to establish some criteria for anomaly calculations for Weyl fermions in order to prevent an annoying increase of entropy and confusion in a subject which, it must be said, is objectively subtle.  To this end we propose a necessary test, which must hold true for any method used to compute anomalies. It consists in checking that the utilized method must be able to reproduce  \textbf{all} the anomalies of a Weyl fermion (not only the trace ones). 
The readers will decide by themselves whether the methods used in \cite{Bast1, Bast2,FZ} 
are fit or not to capture, in particular, the consistent non-Abelian chiral gauge anomaly, the existence of which is definitely uncontroversial, together with the parity odd trace anomaly, whose derivation is similar.

The present comment is organized as follows. First we analyze in detail 
the methods used in the above references and single out criticalities whenever they occur (to be fair, not only in these papers, for they are not infrequent also elsewhere in the literature).  After this we propose our test and explain why some methods are unable to comply with it. Throughout the paper we will keep the discussion at a less technical level as possible. An appendix is devoted in particular to a discussion of the Pauli-Villars regularization.

\subsection*{ Weyl and/or Dirac?}

Let us start with some notations and basic facts. We denote by $\slashed D$ the 
Dirac operator proper: namely,  the massless  matrix-valued differential 
operator applied in general to Dirac spinors on the 4d curved space with 
Minkowski signature $ (+,-,-,-) $
\be
\slashed D= i \left(\slashed \partial + \slashed V\right)\label{dirac}
\ee 
where $V_\mu$ is any real vector potential, including a spin connection in the 
presence of a non-trivial background metric. We use here the four component 
formalism for fermions. The functional integral, i.e. the effective action for a 
quantum Dirac spinor in the presence of a classical background vector potential 
\be
{\cal Z}[V]=\int {\cal D}\psi {\cal D}\bar \psi\; e^{i\int d^4x \, \sqrt{g}\, 
\bar \psi \slashed D \psi}\label{pathint}
\ee
is \textbf{formally } understood as the determinant of $\slashed D\, 
:$ 
$ 
\det \,(\slashed D) = \prod^\infty_i \lambda_i\,.
$
From the concrete and substantial point of view, the latter can be operatively 
defined in two alternative ways: either in perturbation theory, 
i.e. as the sum of an infinite number of 1-loop Feynman diagrams, some of which 
containing UV divergences by naive power counting, 
or in a non-perturbative approach,
i.e. as the suitably regularized infinite product of the eigenvalues of 
$\slashed D$ by means of the analytic continuation tool. 
It is worthwhile to remark that, on the one hand,  the perturbative approach 
requires some UV regulator, in order to give a meaning to a
finite number of UV divergent 1-loop diagrams by naive power counting. On the 
the other hand, in the non-perturbative framework the
complex power construction and the analytic continuation tool, if available,  
both provide by themselves the necessary to set up for the 
infinite product of eigenvalues of a normal operator, without need of any 
further regulator. 

A variation of \eqref{pathint} requires the existence of an inverse of the kinetic operator.
It turns out that an inverse of $\slashed D$ does exist and, if full causality 
is required in forwards and backwards  time evolution on e.g. Minkowski space,
 it is the Feynman propagator or Schwinger distribution $\slashed S$, which is 
unique and characterized by the well-known Feynman prescription, 
 in such a manner that
\be 
\slashed D_x \slashed S(x-y)= \delta(x-y),\quad\quad \slashed D \slashed 
S=1,\label{DP}
\ee
The latter is a shortcut operator notation, which we are going to use 
throughout\footnote{For simplicity we understand factors of $\sqrt{g}$, which 
are necessary, see \cite{DeWitt}, but inessential in this discussion.}.

\vskip 1cm 

The scheme to extract the trace anomaly from the functional integral is 
well-known. It is its response under a Weyl 
(or even a scale) transform $\delta_\omega g_{\mu\nu} = 2 \omega 
g_{\mu\nu}\,:$
\be
 \delta_\omega \log {\cal Z} = \int d^4x\, \omega(x)\, g_{\mu\nu}(x)\, 
\langle\,T^{\mu\nu}(x)\rangle\label{deltaomega}
\ee
where $g_{\mu\nu}(x) \langle T^{\mu\nu}(x)\rangle$ is the quantum trace of the 
energy-momentum tensor. Again, 
the latter can be calculated in various ways with perturbative or 
non-perturbative methods. 
The most frequently used ones are the Feynman diagram technique and the 
so-called analytic functional method, respectively. 
The latter denomination actually includes a collection of approaches, ranging 
from the Schwinger's proper-time method 
\cite{Schwinger}  to the heat kernel method \cite{KlausDima}, the Seeley-DeWitt 
\cite{DeWitt, Seeley} and the zeta-function regularization
\cite{Hawking}.  There is a large consensus on the results concerning the trace 
anomaly for Dirac (and Majorana) fermions and no  significant controversy.
Let us pause to recall, however, that the rigorous and successful implementation 
of both methods and technicalities relies on the possibility to turn to the Euclidean 
formulation, which is just allowed thanks to the above mentioned Feynman 
prescription to get the Schwinger causal Green's function or Feynman propagator 
for the quantum Dirac spinor field. It turns out that, in the absence of such 
a transition to the Euclidean formulation, the whole construction becomes 
mathematically meaningless and unreliable.

When one comes to Weyl fermions things drastically change. Let us denote a 
left-handed Weyl fermion by  $\psi_L= P_L \psi$, where $P_L= \frac12(
{1-\gamma_5})$. Then, for example, the classical action on the 4d Minkowski space reads
\be
S_L=\int \mathrm{d}^4x\, \bar \psi_L \slashed D \psi_L \label{weylaction}
\ee
The Dirac operator, acting on left-handed spinors maps them to right-handed 
ones. Hence,
the Sturm-Liouville or eigenvalue problem itself is not even well posed, so that 
the Weyl determinant cannot even be defined at all. 
This is reflected in the fact that the inverse of
\be
{\slashed D}_L = {\slashed D}P_L = P_R {\slashed D}\label{DL}
\ee
does not exist, since it is the product of an invertible operator times a 
projector. As well, the full propagator of a Weyl fermion does not exist in this 
naive form (this problem can be circumvented in a more sophisticated approach, 
see below). It is incorrect to pretend that the propagator is  $\slashed 
S_L=\slashed S P_R = P_L \slashed S$. First because such an inverse does not 
exist, second because, even formally,
\be
{\slashed D}_L \slashed S_L= P_R,\quad\quad {\rm and}\quad\quad{\slashed S}_L 
{\stackrel {\leftarrow}{ \slashed D_L}}= P_L \label{wrong}
\ee

The inverse of the Weyl kinetic operator is not the inverse of the Dirac operator multiplied by a chiral projector.  Therefore the propagator for a Weyl fermion is not the Feynman propagator for a Dirac fermion multiplied by the same projector.  The lack of an inverse for the chiral Weyl 
kinetic term has drastic consequences even at the free non-interacting level. For instance, the evaluation of the functional integral 
(i.e. formally integrating out the spinor fields) 
does involve the inverse of the kinetic operator: thus, it is clear that the corresponding formulas for the chiral Weyl quantum theory 
cannot exist at all, 
so that no Weyl effective action can be actually defined even in the free non-interacting case.
Let us add that considering the square of the kinetic operator (see below), as it is often done in the literature, does not change the conclusion.

It may sound strange that the (naive) full propagator for Weyl 
fermions does not exist, especially if one has in mind perturbation 
theory in the Minkowski space. In that case, in order to construct Feynman diagrams, one uses the ordinary {\bf free}
Feynman propagator for Dirac fermions. The reason one can do so is because the 
information about chirality is preserved by the fermion-boson-fermion vertex, 
which contain the $P_L$ projector (the use of a free Dirac propagator is 
formally justified, because one can add a  free right-handed fermion 
to allow the inversion of the kinetic operator, see the Appendix below).
On the contrary, the full propagator is supposed to contain the full chiral 
information, including the information contained in the vertex, i.e. the 
potentials. If one  pretends to replace the full Weyl propagator with the full 
Dirac propagator, one loses  any information concerning the chirality, and 
trying to recover it by multiplying the Dirac propagator by $P_L$ is far too 
naive (see below). 
\vskip 1cm

After these preparatory remarks, let us analyze the first critical key issue, not uncommon in the literature, but which in \cite{FZ} is applied to the calculation of trace anomalies. The authors, following 
DeWitt \cite{DeWitt} and Christensen \cite{chris}, in order to evaluate the trace anomaly 
of a Dirac fermion use the proper time method, splitting the point where the two fermion fields in $T_{\mu\nu}$ sit, and using the (full) Dirac propagator (the Hadamard 
function is twice the imaginary part of the latter). The only modification they introduce for Weyl fermions consists  in inserting a chiral projector multiplying a regularized split e.m. tensor appropriate for Dirac fermions. As we have just explained this is wrong in principle; but it 
is useful to delve a bit into these (erroneous) formulas in order to better 
appreciate the origin of the (wrong) null result for the split trace obtained by the authors. 
The point is that the quantization process (via point splitting) is the one 
appropriate for a Dirac fermion and only at the end is the result evaluated 
(tracing over gamma matrices) after multiplying the quantum expression valid for 
a Dirac fermion by a chiral projector. It is rather obvious that in so doing one 
gets a vanishing odd-parity part of the trace, because a Dirac fermion contains 
both chiralities and, being left-right symmetric, 
cannot produce any left-right asymmetry. So, eventually inserting a 
chiral projector cannot resuscitate what has been eliminated from the beginning! The  contribution proportional to $\gamma_5$ inevitably vanishes. We repeat once more: 
respecting the appropriate chirality throughout the calculation is essential when Weyl fermions are involved, otherwise whatever calculation may lead to fake results. A crucial test 
is the chiral consistent anomaly \eqref{A} (see below), which is impossible 
to reproduce by the just described  method of ref. \cite{FZ}\footnote{Notice that the very same obstacles and obstructions occur  even in the two-component form of the Weyl 
spinors. {Rewriting the 4d formulas in the two-component formalism cannot change the result.} The crux is that, in the two-component formalism, the kinetic operator for Weyl fermions is the same as the kinetic operator for Majorana fermions (see the relevant discussion below). Since in the non-perturbative approach fermions are integrated out, all the information is stored in the kinetic operator. The consequence is that any notion of chirality is lost. It is not a surprise that the results obtained in this way are appropriate for Majorana  fermions, but not for Weyl fermions. The two-component formalism for Weyl fermions, although sometime used in the literature, is not reliable when chirality is a crucial issue.}.  {This method can only reproduce covariant (non-split) anomalies.}

\subsection*{Weyl and/or Majorana?}

The second criticality is hidden in the (wrong) identification between Weyl and Majorana massless spinors. 
This is not an uncommon attitude in the literature. It is also the basis of \cite{Bast1,Bast2}, where a Fujikawa-like method, \cite{Fujikawa}, is employed to compute anomalies. In their approach several aspects deserve a close inspection.

Let us start by remarking that the Fujikawa method is a simplified heuristic version of the so called 
heat-kernel expansion method.general 
 It is well-known since long, \cite{AB}, that this method is bound to run into difficulties when Weyl fermions are involved. 
 Its underlying idea is to compute the variation of the fermion measure. 
 However, as we have pointed out, this proposal does not make sense for a (naive) Weyl-Dirac operator. In order to circumvent this difficulty one may try to resort to some elliptic 
operator (for a Dirac fermion it is, e.g., the square of the Dirac operator). 
Now, it is not difficult to concoct an elliptic operator out of the Weyl-Dirac one, 
but it is impossible to do it while preserving all the critical information, 
i.e. while preserving the chirality of the model and its classical symmetries.

There are in fact insurmountable difficulties. As a matter of fact, first of all, in order to turn the square of the Dirac operator  
into an elliptic normal differential and matrix-valued operators,  the transition to the 
Euclidean formulation is mandatory. The latter is absolutely legitimate and 
viable for Dirac fermions, whilst it does not exist at all for Weyl fermions. 
The very reason is deep and sharp: the Euclidean 4d symmetry group is 
$ O(4,\mathbb{R}) $ which is locally isomorphic to the direct  product $ O(3,\mathbb{R})\times O(3,\mathbb{R}) $, in such a manner that any item in this theory must be invariant with respect to the exchange of any spin 
representation of the two identical and equivalent orthogonal trivial factors, no room being left 
to the very concept of chirality, which requires two non-equivalent irreducible 2d 
representations of the Lorentz group.

Another difficulty comes from the construction itself of \cite{Bast1,Bast2}.
The claim of the authors  is that a Weyl fermion is the same as a massless 
Majorana fermion. In keeping with this idea, they use a differential operator appropriate for a 
massless Majorana fermion. The end result is that they add to the chosen operator its charge conjugate. Enough is to say that charge conjugation maximally violates chirality, since, 
for instance,
\be
{\EC}\left( i \overline{\psi_L }\gamma^\mu \partial_\mu
\psi_L\right)  {\EC}^{-1}
= i \overline{\psi_R }\gamma^\mu \partial_\mu
\psi_R.\label{CactionLH}
\ee
Moreover, this addition ends up breaking some classical symmetry of the 
theory\footnote{Mathematically speaking, the addition to an operator of  its 
charge conjugate one is normally a nonsense,  unless they have the same domain and codomain.}. 
It is clear that within such a scheme it is impossible to preserve chirality and intercept any split anomaly.

In \cite{Bast1,Bast2} there is also a potential  problem of regularization 
dependence. In non-perturbative approaches, as already explained,
the effective action can be properly and rigorously defined iff the differential 
operator, which appears in the classical action and specifies its symmetries,
does satisfy all the hypotheses which are required by the Seeley theorems 
\cite{Seeley}. In such a circumstance, the effective action is properly and
rigorously defined by the zeta-function regularization and its variation with 
respect to the classical symmetry groups is as well defined, without any need of any 
further regulator. But it is well-known that this procedure cannot be 
applied to the Weyl-Dirac operators, just owing to chirality, as we have just 
explained.

In the absence of a solid and reliable definition of the Weyl effective 
action, the authors of \cite{Bast1,Bast2} use a mixture of non-perturbative Fujikawa heuristic method and old-fashioned proper-time
 or Schwinger-DeWitt method, together with a {  regularization procedure they call Pauli-Villars, 
although it does not have anything in common with the original PV method to regulate Feynman diagrams, but for the presence of a mass. 
As a preliminary remark, we notice that } 
the  old-fashioned proper-time method is  mathematically non-rigorous and has been thoroughly improved 
along the years and replaced by the modern Seeley-DeWitt approach supplemented by the Hawking method 
of the zeta-function regularization to get a rigorous non-perturbative definition of the functional determinants. 
All this is well-known. In their {miscellaneous  approach, } the authors of \cite{Bast1,Bast2} use the above mentioned square of the 
Dirac-Majorana operator as a starting point together with what they call Pauli-Villars one loop effective action  to treat divergences.  
Now, first of all the quadratic operator inevitably breaks conservation of chirality, since it involves both chiralities in a balanced form 
and excludes any possibility to capture split anomalies.  Moreover 
the perturbative one loop corrections to the classical Weyl action, as obtained from the original true Pauli-Villars regularization method, 
can by no means generate any dynamical mass term, neither Dirac nor Majorana, since a PV regularization preserves gauge invariance 
and U(1) internal chiral phase transformation invariance, so that it protects classical scale invariance and forbids any mass term at the quantum level. 
As a consequence,  it is rather misleading to call Pauli-Villars the would be one loop modified effective Lagrangian of \cite{Bast1,Bast2}. 
The true PV regularization in the perturbative Feynman diagram is illustrated as an example in Appendix, 
where it is shown that it dramatically differs from the regularization proposed in \cite{Bast1,Bast2}.

Concerning general regularization methods involving spinor mass terms, just like the one proposed in \cite{Bast1,Bast2} 
and improperly called Pauli-Villars \cite{PV},
 it is in principle not a priori fit in a calculation where the preservation of chirality is crucial, since it introduces both chiralities on an equal footing. Whatever results one obtains with that method in such context, they should be crosschecked with other regularizations. As a matter of
 fact, the perturbative evaluation of the one loop effective action, for a Weyl spinor minimally coupled to a gauge vector potential, yields
 essentially the same universal massless result, no matter the employed regulators (see Appendix). 

To sum up, the use of a Majorana-Dirac operator inevitably drives to a chirally symmetric operator, which cannot produce any left-right asymmetry, i.e. no chirally split anomaly. This does not affect simply the Pontryagin anomaly. For the same reason this method, barring radical changes, is unable to intercept the consistent chiral anomaly \eqref{A} below.

\section*{Dirac-Weyl-Majorana}

It is not beside the point to insist again on the distinction between Weyl and 
massless Majorana fermions. This issue has been discussed also elsewhere, see for instance 
\cite{BCDDGS}. Here we briefly return to the main points. Massive and massless 
Majorana spinors have really nothing to share with chiral Weyl spinors.  A 
classical Majorana spinor 
is a self-conjugated bispinor, that can be always chosen to be real and always 
contains both chiralities in terms of four real 
independent component functions. It describes neutral spin 1/2 objects - not yet 
detected in Nature - and consequently there is
no phase transformation (U(1) continuous symmetry) involving self-conjugated 
Majorana spinors, independently of the presence or not of a mass term. Hence, 
e.g., its particle states do not admit antiparticles of opposite charge, simply 
because charge does not exist at all for charge self-conjugated
spinors (actually, this was the surprising discovery of Ettore Majorana, 
after the appearance of the Dirac equation and the positron detection).
The general solution of the wave field equations for a free Majorana spinor 
always entails the presence of  two polarization states with opposite helicity.
On the contrary, it is well known that a chiral Weyl spinor, describing massless
neutrinos in the Standard Model, admits only one polarization or helicity state,
it always involves antiparticles of opposite chirality and helicity, it always 
carries a conserved internal quantum number such as the lepton number.

At this point it is perhaps not useless to clarify an issue concerning the just mentioned U(1) continuous symmetry of Weyl fermions. The latter is often confused with an axial ${\mathbb R}$ symmetry of Majorana fermions and used to justify the identifiction of Weyl and massless Majorana fermions. To start with let us consider a free massless Dirac fermion $\psi$. Its free action is clearly invariant under the transformation $\delta\psi =i (\alpha +\gamma_5\beta)\psi$, where $\alpha$ and $\beta$ are real numbers. This symmetry can be gauged by minimally coupling $\psi$ to a vector potential $V_\mu$ and an axial potential $A_\mu$, in the combination $V\mu +\gamma_5 A_\mu$, so that $\alpha$ and $\beta$ become arbitrary real functions. For convenience let us choose the Majorana representation for gamma matrices, so that all of them, including $\gamma_5$, are imaginary. If we now impose $\psi$ to be a Majorana fermion, its four component will be real and only the symmetry parametrized by $\beta$ makes sense in the action (let us call it $\beta$ symmetry). If instead we impose $\psi$ to be Weyl, say $\psi=\psi_L$, then , since $\gamma_5 \psi_L=\psi_L$, the symmetry transformation will be  $\delta\psi_L =i (\alpha -\beta)\psi_L$.

We believe this may be the origin of the confusion, because it looks like we can merge the two parameters $\alpha$ and $\beta$ into a unique one and identify it with the $\beta$ of the Majorana axial $\beta$ symmetry. However this is not so because for a  right handed Weyl fermion  the symmetry transformation is $\delta\psi_R =i (\alpha + \beta)\psi_R$. Forgetting $\beta$,
the Majorana fermion does not transform. Forgetting $\alpha$, both Weyl and Majorana fermions transform, but the Weyl fermions transform with opposite signs for opposite chiralities. 
In terms of anomalies, it is well-known that the axial ${\mathbb R}$ Majorana symmetry is anomalous: this is the well-known covariant anomaly $\sim \int d^4x \, \beta F_A \wedge F_A$ ($F_A$ is the curvature of $A$). For a left (right) Weyl fermion the symmetry with parameter $\alpha-\beta$ ($\alpha+\beta$) is anomalous. This is a consistent anomaly, which, in the Abelian case we are considering, coincides in form with the covariant anomaly (but not in the non-Abelian case, see below), although with a different coefficient and with opposite signs for opposite chiralities. Since a Dirac fermion can be regarded as a sum of two Weyl fermions with opposite chiralities, we see that the anomalies triggered by the $\alpha$ transformation cancel out, while the anomalies triggered by $\beta$ add up. This is consistent with the well-known fact that for a Dirac fermion the  $U(1)$ $\alpha$ symmetry is not anomalous, while the axial $\beta$ symmetry is anomalous and corresponds to twice the anomaly of   a Majorana fermion. As we see the symmetries and anomalies of Majorana fermions are different from the symmetries and anomalies of Weyl fermions.

The previous discussion is qualitatively correct but it does not take care completely of the renormalization of the anomaly coefficient of a Dirac fermion when considered as the sum of two Weyl fermions. The complete treatment is based on the Bardeen method and is outlined below.

In  conclusion, Weyl and massless Majorana spinors possess quite opposite
physical properties, the latter being much closer, if eventually detected, 
 to those of a real photon (no charge, no parity breaking,  two polarization 
states). Thus, it is absolutely evident that Weyl and massless Majorana spinors can by no 
means  be identified. In particular, in this anomaly problem where chirality is 
fundamental, such identification is fully misleading.

\subsection*{A necessary test}

Next we lay out in detail the above advocated necessary
consistency test.
Let it be clear, first of all, that checks of non-trivial results are necessary 
and welcome. However, they should be made on an unambiguously homogeneous 
footing. In this kind of trade there are rules that must be respected. One of 
the indisputable rules is that a unique method should be used to calculate 
the anomalies of a given theory.
If the methods used in refs.\cite{Bast1, Bast2,FZ} lead to a vanishing 
odd-parity trace anomaly, they should be able, on the other hand, to compute at 
least all the other known anomalies of a Weyl fermion. If they are not, it means 
that such methods are unreliable (however we hasten to add that, if they are,  
it does not mean by itself that the methods are reliable: another condition is that also the trace 
anomalies must be consistently dealt with along the same lines).

A very important anomaly is the consistent chiral anomaly, which is produced 
when a Weyl fermion couples to a gauge potential $A_\mu= A_\mu^a T^a$, where 
$T^a$ are Lie algebra generators. We write down here its explicit expression in 
flat background to avoid any misunderstanding\footnote{We are not referring to the 
covariant anomaly, which exists  also for Dirac and Majorana fermions, but to 
the consistent one, and, specifically, to the non-Abelian anomaly, because 
Abelian covariant and consistent gauge anomalies accidentally coincide in form 
(although their coefficients are different), see Appendix for an example.} 
\be
 {\cal A}(\lambda)=\pm \frac 1{24\pi^2} \int d^4x\, \epsilon^{\mu\nu\lambda 
\rho}\, \tr\left( \lambda\, \partial_\mu\left( A_\nu\partial_\lambda A_\rho+ 
\frac 12 A_\nu A_\lambda A_\rho\right)\right) \label{A}
\ee
where $\lambda=\lambda^a T^a$ is the parameter of the ordinary infinitesimal 
gauge transformation, namely,
\be
\delta_\lambda A_\mu = \partial_\mu \lambda +[A_\mu, \lambda] \label{gaugetr}
\ee
As above the $\pm$ sign depends on the chirality. This is another example of 
chirally split anomaly.

The anomaly \eqref{A} is the response  under \eqref{gaugetr} of 
$$ {\cal Z}_L[A] =\int {\cal D}\psi {\cal D}\bar \psi\:e^{i\int d^4x \, \bar 
\psi_L \slashed D \psi_L}$$
i.e.  the functional integral for a Weyl fermion multiplet coupled to a 
non-Abelian gauge field. 
This anomaly has been computed in many different ways, and a correct method to 
evaluate any quantity from the path integral  $ {\cal Z}_L[A] $  must be able to 
reproduce it. Now, if we submit the methods used in \cite{Bast1, Bast2,FZ} to 
this crucial test one can see immediately, and without doing any calculation, 
that, unless one introduces drastic changes in them, the anomaly \eqref{A} will 
never be produced. The reason is that such methods do not respect the chiral 
splitting. For different reasons and with different approaches they are bound to a chirally 
symmetric description. 

\section*{Conclusion}

In conclusion, the two critical issues illustrated above put in jeopardy the results for anomaly computations for Weyl fermions. In order to find a correct procedure in this kind of problems one has to employ somewhat more sophisticated methods.

The consistent anomaly \eqref{A} can be captured
either by a perturbative or a nonperturbative method, using the
Bardeen approach, \cite{Bardeen}. W. A. Bardeen used Dirac fermions coupled to a vector $V$ and 
an axial potential $A$. Using a point splitting regularization he eventually considered the 
limit $V\to A/2$ and $A\to A/2$, obtaining in this way \eqref{A}. Actually this limit is completely smooth, so that one can take it at the very beginning. This implies  starting, for a left-handed fermion, from the  kinetic operator
\be
i \gamma^\mu \left(\partial_\mu + P_L V_\mu\right), \label{bardeen}
\ee
which is invertible  and in accord with the above mentioned Feynman diagram approach.  This is the procedure used by B. DeWitt in chapter 28 of his book, \cite{DeWittglobal}, to obtain again \eqref{A}. It is clear that in order to intercept the split trace anomaly for Weyl fermions one must proceed along the same lines, i.e. consider, together with a metric, a companion axial potential.

This method has been explored in \cite{BGL,BDL,BCDDGS}. Admittedly, although the perturbative method provides evidence for the existence of the odd-parity trace anomaly, in order to 
obtain decisive results, one should push the approximation further on and, for 
instance, compute at least four-point amplitudes. An alternative way is the 
non-perturbative method  of ref. \cite{BCDGPS}.   It mimics the just illustrated Bardeen approach, as implemented by B. DeWitt, \cite{Bardeen, DeWittglobal}, to derive the consistent gauge anomaly. It is  a heat kernel method  adapted to a hypercomplex analytical framework. It bypasses the difficulty of dealing with the Weyl-Dirac operator by introducing an axial symmetric potential 
beside the usual metric in a theory of Dirac fermions, and eventually taking an 
appropriate limit to recover a Weyl fermion setting. The relevant heat kernel 
elliptic operator is obtained starting from the physical principle that it must preserve 
all the symmetries of the classical theory. The results are obtained using two 
different regularizations: the dimensional and zeta function regularizations. 
The method is fully covariant, produces a nontrivial odd-parity trace anomaly 
and (as was shown fifty years ago by W.A. Bardeen) yields the consistent, 
\eqref{A}, as well as the covariant gauge anomalies. So far this is the only 
successful and thorough method. 

There hopefully exist other variants. Any new calculation and verification in 
this context is most welcome,  but in order to seriously challenge the results 
of \cite{BCDGPS}, one must envisage methods that preserve the chirality as well 
as all the symmetries of the classical theory in the choice of the elliptic operator. 
Moreover, they must satisfy the 
minimal condition of producing all the already known anomalies  for a Weyl 
fermion and, in particular, \eqref{A}. And they must derive the trace anomalies along the same lines. Without passing this test any result for the trace anomaly of Weyl 
fermions cannot be homologated. 

To conclude our comment we consider two recent papers \cite{Bast3,Bast4} and analyze whether they respect the just enunciated criteria. In \cite{Bast3} the authors apply the Bardeen method (adding an axial potential, beside the vector one, in a model of Dirac fermions) to compute the consistent gauge anomaly for Weyl fermions. They succeed, but details are important: they use a Fujikawa method (with the relevant quadratic operator for Dirac fermions) and  {\it what they call a PV regularization in which the massive ghosts are Dirac fermions}.
 This part of \cite{Bast3} leads to correct results for consistent gauge anomalies.
But then, on the basis of this calculation, the authors pretend to prove  that the trace anomaly of Weyl fermions coupled to a gauge potential does not contain the odd parity partner. We simply notice that, just as in the gauge case, it is not possible to reproduce the consistent anomaly \eqref{A} without introducing an axial partner, in fact it is not possible to deduce anything about the trace anomaly in a nontrivial background without exciting an axial partner to the metric. Had the authors followed this more complete approach (the analog as for the consistent gauge anomaly), they would have found a nontrivial odd parity gauge contribution to the trace anomaly.  In a more recent paper the same authors \cite{Bast4} finally decide to introduce an axial partner to the metric. They apply the usual heuristic method to determine a quadratic operator and use {their} so-called PV regularization. Their end result is that they do not find the Pontryagin anomaly. However, once again, details matter. Surprisingly, and contradicting \cite{Bast3}, they do not use a Dirac mass regularization, {\it but a Majorana mass one}. This means that their quadratic operator is self-conjugate. In particular it violates a significant part of the symmetries of the model (the authors explicitly disregard the axial diffeomorphisms).  We know, from the above discussion that in this case there is no hope to intercept any split anomaly. In conclusion, neither \cite{Bast3} nor \cite{Bast4} respect the above established consistency criteria. 

\vskip 2cm
\section*{Appendix. The true PV regularization}

The purpose of this Appendix is to show the differences between the Pauli-Villars regularization in  perturbative field theory and the regularization used in \cite{Bast1,Bast2}, which is improperly referred to as PV regularization. To this end we analyze the elementary model of a Weyl fermion coupled to an electromagnetic field.

\medskip
The classical Lagrange density for a Weyl (left) spinor in the four component formalism 
\[
\psi(x)=\chi_{L}(x)=\left\lgroup
\begin{array}{c}
\chi(x) \\ 0
\end{array}
\right\rgroup
\]
reads
\[
\mathcal{K}(x)=\overline{\psi}(x)\,i\partial\!\!\!/\,\psi(x)=\chi_{L}^{\dagger}(x)\,\alpha^{\nu}i\partial_{\nu}\chi_{L}(x), 
\quad\quad \gamma^{\,\nu} = \gamma^0 \alpha^{\nu}
\]
It follows that the corresponding matrix valued Weyl differential operator
\[
w_{L}\equiv\alpha^{\nu}i\partial_{\nu}P_{L},\qquad\quad
P_{L}=\textstyle\frac12(\mathbb{I}-\gamma_5)
\]
is singular and does not possess any rank-four inverse. After minimally coupling to a real massless vector field $ A^{\mu}(x) $
we come to the classical Lagrangian
\[
\mathcal{L}=\chi_{L}^{\dagger}\,\alpha^{\nu}i\partial_{\nu}\chi_{L}
+ gA^{\nu}\,\chi_{L}^{\dagger}\,\alpha_{\nu}\chi_{L} -\textstyle\frac14\,F^{\,\mu\nu}\,F_{\mu\nu}
\]
where $ 0\le g< 1 $ is a non-negative small parameter, while $ F_{\mu\nu}=\partial_{\mu}A_{\nu}-\partial_{\nu}A_{\mu}\,. $
It turns out that the classical action 
\[
S=\int\mathrm{d}^{4}x\,\mathcal{L}
\]
is invariant under the Poincar\'{e} group,  as well as under  the internal U(1) phase transformations
$ \chi_L(x)\mapsto\,e^{\,ig\theta}\chi_{L}(x)\,. $ It is also invariant under the so called scale or
dilatation transformations, viz.,
\begin{eqnarray*}
x^{\,\prime\mu}=e^{\varrho}x^{\,\mu}\qquad \chi_L^{\,\prime}(x)=e^{\,\frac32\varrho}\chi_L(e^{\,\varrho}x)
\qquad A^{\prime\mu}(x)=e^{\,\varrho}A^{\mu}(e^{\,\varrho}x)
\end{eqnarray*}
with $ \varrho\in\mathbb{R} $, as well as with respect to the local phase or gauge transformations
\[
\chi^{\,\prime}_L(x)=\,e^{\,ig\theta(x)}\chi_{L}(x)\qquad\quad A^{\prime}_{\nu}(x)=A_{\nu}(x)+\partial_{\nu}\theta(x)
\]
which amounts to the ordinary U(1) phase transform in the limit of constant phase. It follows therefrom that there are twelve conserved charges
in this model at the classical level and, in particular, owing to scale and gauge invariance, no mass term is allowed for both spinor and vector
fields. The question naturally arises if those symmetries hold true after the transition to the quantum theory and, in particular, if they are
protected against loop radiative corrections within the perturbative approach. Now, in order to develop perturbation theory, one has to face
the problem of the lack of an inverse kinetic operator for both the Weyl and gauge fields, owing to chirality and gauge invariance. In order to solve this problem, it is 
expedient to add to the Lagrangian non-interacting terms, which are fully decoupled from any physical quantity. 
\textbf{These terms break  chirality and
gauge invariance}, albeit in a harmless way, just to allow the setting up of a Feynman propagator, or causal Green's functions,
for both the Weyl and gauge quantum fields. The simplest choice, which preserves 
Poincar\'{e} and internal U(1) phase change symmetries,
is provided by (see \eqref{bardeen})
\[
\mathcal{L}^{\,\prime}=\varphi_{R}^{\dagger}\,\alpha^{\nu}i\partial_{\nu}\varphi_{R} 
-\textstyle\frac12(\partial\cdot A)^{2}
\]
where 
\[
 \varphi_{R}(x)=\left\lgroup
\begin{array}{c}
0 \\ \varphi(x)
\end{array}
\right\rgroup
\]
is a left-chirality breaking right-handed Weyl spinor field. Notice \textit{en passant} that the modified Lagrangian 
$ \mathcal{L}+\mathcal{L}^{\prime} $ does exhibit the futher U(1) internal symmetry under the so called chiral phase transformations
\[
\psi^{\prime}(x)=(\cos\theta+ig\sin\theta\,\gamma_{5})\psi(x)
\qquad\quad\psi(x)=\left\lgroup
\begin{array}{c}
\chi(x) \\ \varphi(x)
\end{array}
\right\rgroup
\]
so that the modified theory involves another conserved charge at the classical level.
From the modified Lagrange density we get the Feynman propagators for the massless Dirac field $ \psi(x) $,
as well as for the massless vector field in the so called Feynman gauge: namely,
\[
S(p)=\frac{ip\!\!/}{p^{2}+i\varepsilon}\qquad\quad D_{\mu\nu}(k)=\frac{-ig_{\mu\nu}}{k^{2}+i\varepsilon}
\]
and the vertex $ ig\gamma^{\nu}P_L\,,  $ with $ k+p-q=0 $, which involves a vector particle of momentum $ k $
and a Weyl pair of particle and anti-particle of momenta $ p $ and $ q $ respectively and of opposite helicity.\footnote{Customarily, 
the on-shell 1-partiche states of a left Weyl spinor field are a left-handed particle with negative helicity 
$ -\frac12\hslash $ and  a right-handed antiparticle of positive helicity $ \frac12\hslash $.}

The lowest order 1-loop correction to the Weyl kinetic term $ p\!\!/P_L $ is formally provided by the Feynman rules
in the Minkowski space, viz.,
\[
\Sigma_2(p\!\!/) =
 -\,ig^{\,2}\int\frac{\mathrm{d}^{4}\ell}{(2\pi)^{4}}\;
\gamma^{\,\mu}\,D_{\mu\nu}(\,p-\ell\,)\,S(\ell)\,\gamma^{\,\nu}P_L
\]
By naive power counting the above 1-loop integral turns out to be UV divergent. Hence a regularization procedure is
mandatory to give a meaning and evaluate the radiative correction $ \Sigma_2(p\!\!/) $ to the Weyl kinetic operator. 
In the sequel we shall examine in detail the Pauli-Villars (PV) regularisation and comment about the most popular alternatives,
such as the dimensional (DR) and UV cut-off (UV) regularizations.

\medskip
\textsc{Pauli-Villars regularization}

\medskip
Let us now perform the calculation for the left Weyl spinor self-energy in the Pauli-Villars regularization.
The latter is simply implemented by the following replacement of the massless Dirac propagator
\[
\mathtt{reg}\,\Sigma_2(p\!\!/) =
 -\,ig^{\,2}\int\frac{\mathrm{d}^{4}\ell}{(2\pi)^{4}}\;
\gamma^{\,\mu}\,D_{\mu\nu}(\,p-\ell\,)\,
\sum_{s\,=\,0}^S C_s\,S(\ell\,,\,M_s)\,
\,\gamma^{\,\nu}P_L
\]
where $M_0 = 0\,,\ C_0=1$ while 
$\{\,M_s\equiv \lambda_s\,M\,|\,\lambda_s\gg 1\ (\, s=1,2,\ldots,S\,)\,\}$
is a collection of very large auxiliary masses. The set of constants
$C_s$ will be suitably selected,
as we shall see in the sequel, in such a manner as
to obtain a specific and mathematically meaningful form for the 
ultraviolet divergences that will manifest themselves in the limit
$\lambda_{s} \rightarrow\infty\,.$
Since we have
\[
S(\ell\,,\,M_s)=\frac{ i(\ell\!\!/ + M_{s}) }{ \ell^{\,2} - M_{s}^{2} + i\varepsilon }
\qquad\quad(\,s=1,2,\ldots,S\,)
\]
it is now convenient to set
\[
\mathtt{reg}\,\Sigma_2(p\!\!/) \equiv f(p^{\,2})\,p\!\!/P_L - M\upsilon(\,p^{\,2})
\]
in such a manner that we can write
\begin{eqnarray*}
\mathrm{tr} \left[ \,\mathtt{reg}\,\Sigma_2(p\!\!/)\,\right]  &=& -\,4M\upsilon(\,p^{\,2})
\\
&=& -\,ig^{\,2}\int\frac{\mathrm{d}^{4}\ell}{(2\pi)^{4}}\;\sum_{s\,=\,1}^S C_s\,
\frac{ 8M_s }{ [\,(\ell-p)^{2}+i\varepsilon\,] \left( \ell^{\,2} - M_{s}^{2} + i\varepsilon \right)  }
\\
\mathrm{tr} \left[ \,p\!\!/\mathtt{reg}\,\Sigma_2(p\!\!/)\,\right]  &=& 2p^{\,2}f(\,p^{\,2})
\\
&=& ig^{\,2}\int\frac{\mathrm{d}^{4}\ell}{(2\pi)^{4}}\;\sum_{s\,=\,0}^S C_s\,
\frac{ 4\,p\cdot\ell }{ [\,(\ell-p)^{2}+i\varepsilon\,] \left( \ell^{\,2} - M_{s}^{2} + i\varepsilon \right)  }
\end{eqnarray*}
Once again, if we take advantage of the Feynman parametric formula we obtain
\begin{eqnarray*}
\upsilon(\,p^{\,2})&=&
\int\frac{\mathrm{d}^{4}\ell}{(2\pi)^{4}}\;\sum_{s\,=\,1}^S C_s\int_{0}^{1}\mathrm{d}x\,
\frac{ 2ig^{\,2}\lambda_s }{\left[\,\ell^{\,2} - 2x\,p\cdot \ell + xp^{\,2}  - (1-x)M_{s}^{2} + i\varepsilon\,\right]^2}
\\
&=&\int_{0}^{1}\mathrm{d}x\int\frac{\mathrm{d}^{4}\ell^{\,\prime}}{(2\pi)^{4}}\;\sum_{s\,=\,1}^S 
\frac{ 2ig^{\,2}C_s\lambda_s }{\left[\,\ell^{\,\prime 2}  + x(1-x)p^{\,2}  - (1-x)M_{s}^{2} + i\varepsilon\,\right]^2}
\\
&=&-\,2g^{\,2}\int_{0}^{1}\mathrm{d}x\int\frac{\mathrm{d}^{4}\ell_E}{(2\pi)^{4}}\;\sum_{s\,=\,1}^S 
\frac{ C_s \lambda_s }{\left[\,\ell_{E}^{\,2}  + x(1-x)p_{E}^{\,2}  + (1-x)M_{s}^{2} \,\right]^2}
\end{eqnarray*}
In order to proceed further on it is convenient to consider the generating integral
\[
I_{\,n}\,(z_E) \equiv
(-1)^{n}\int\frac{\mathrm{d}^4 \ell_E}{(2\pi)^{4}}
\sum_{s=1}^S C_s\lambda_{s}\,\frac{\exp\lbrace i\ell_E\cdot z_E\rbrace}
{(\ell_E^{\,2}+\Delta_{s}^{2})^{n}}
\]
where $ \Delta_{s}^{2}=(1-x)[\,xp_{E}^{\,2}  + M^{2}\lambda_{s}^{2}\,]>0\,,\ \forall s=1,2,\ldots,S\,. $
Explicit calculation yields
\begin{equation}
I_{\,n}\,(z_E) = \frac{(-1)^{\,n}}{8\pi^2\,\Gamma(n)}\,
\sum_{s=1}^S C_s\lambda_{s}\,\left(\frac{2\,\Delta_{s}}{|\,z_E\,|}\right)^{2-n}\,
K_{\,2-n}(\,\Delta_{ s}\,\vert\,z_E\,\vert\,)
\end{equation}
where $ z_{E}=(\mathbf{z},z_{4}) $ while $ \vert\,z_E\,\vert=\sqrt{\mathbf{z}^{2}+z_4^{2}}\,, $
whence we immediately obtain 
$$
I_{\,2}\,(z_E) =
\frac{1}{8\pi^2}\,
\sum_{s\,=\,1}^S C_s\lambda_{s}\,K_{0}(z_{s})\qquad\quad(\,z_{s}=\Delta_{ s}\,\vert\,z_E\,\vert\,)
$$
where $K_{0}$ is the modified Bessel function, the series representation of which is provided by
$$
K_{0}(z)=\sum_{k\,=\,0}^\infty
\frac{1}{(k!)^2}\left(\frac{z}{2}\right)^{2k}
\left[\,\psi(k+1)-\ln\frac{z}{2}\,\right]=\ln\frac{2}{z} -\,\mathbf{C}+O(z^{2}\ln z)
$$
where \textbf{C} denotes the Euler-Mascheroni constant. Then we can write
\[
\frac{1}{4M}\,\mathrm{tr} \left[ \,\mathtt{reg}\,\Sigma_2(p\!\!/)\,\right] = -\,\upsilon(\,p^{\,2}) = 2g^{\,2}\int_{0}^{1}\mathrm{d}x\,
\lim_{\,z_{E}\to\,0}I_{\,2}\,(z_E) 
\]
whence it is clear that the limit exists, so that the regularization works, iff the Pauli-Villars condition is fulfilled, viz.,
\[
\sum_{s\,=\,1}^S C_s\lambda_{s}=0
\]
which provides in turn $ \upsilon=0 $, i.e., no mass term is allowed in the 1-loop Weyl kinetic term, both
in dimensional  and Pauli-Villars regularizations, as expected.

Let us turn now to the other form factor
\begin{eqnarray*}
&&\mathrm{tr} \left[ \,p\!\!/\mathtt{reg}\,\Sigma_2(p\!\!/)\,\right]  = 2p^{\,2}f(\,p^{\,2})
\\
&=& ig^{\,2}\int\frac{\mathrm{d}^{4}\ell}{(2\pi)^{4}}\;\sum_{s\,=\,0}^S C_s\,
\frac{ 4\,p\cdot\ell }{ [\,(\ell-p)^{2}+i\varepsilon\,] \left( \ell^{\,2} - M_{s}^{2} + i\varepsilon \right)  }
\\
&=&\int\frac{\mathrm{d}^{4}\ell}{(2\pi)^{4}}\;\sum_{s\,=\,0}^S C_s\int_{0}^{1}\mathrm{d}x\,
\frac{ 4ig^{\,2}p\cdot\ell }{\left[\,\ell^{\,2} - 2x\,p\cdot \ell + xp^{\,2}  - (1-x)M_{s}^{2} + i\varepsilon\,\right]^2}
\\
&=& -\,2g^{\,2}p^{\,2}\int_{0}^{1}\mathrm{d}x\,x\int\frac{\mathrm{d}^{4}\ell_E}{(2\pi)^{4}}\;\sum_{s\,=\,0}^S 
\frac{ C_s  }{\left[\,\ell_{E}^{\,2}  + x(1-x)p_{E}^{\,2}  + (1-x)M_{s}^{2} \,\right]^2}
\end{eqnarray*}
whence we obtain
\begin{eqnarray*}
f(p^{\,2}) &=& -\,g^{\,2}\int_{0}^{1}\mathrm{d}x\,x\int\frac{\mathrm{d}^{4}\ell_E}{(2\pi)^{4}}\;\sum_{s\,=\,0}^S 
\frac{ C_s  }{\left[\,\ell_{E}^{\,2}  + \Delta_{s}^{2} \,\right]^2}
\\
&=&-\,\frac{g^{\,2}}{16\pi^{2}}\int_{0}^{1}\mathrm{d}x\,x  
\int_{0}^{\infty}\frac{\mathrm{d}\tau}{\tau}\sum_{s\,=\,0}^S C_s 
\exp \left\lbrace -\,\tau(1-x)[\,xp_{E}^{\,2}  + M^{2}\lambda_{s}^{2}\,]\right\rbrace 
\\
&=&-\,\frac{g^{\,2}}{16\pi^{2}}\int_{0}^{1}\mathrm{d}x\,x  
\int_{0}^{\infty}\frac{\mathrm{d}\tau}{\tau}
\exp \left\lbrace \tau x(1-x)\,p^{\,2}\right\rbrace
\\
&\times&\sum_{s\,=\,0}^S C_s \left[\, 1-\sum_{n=1}^{\infty}\frac{\tau^{\,n}}{n!}(1-x)^{n}M^{2n}\lambda_{s}^{2n}\,\right]
\end{eqnarray*}
Integrability requires the further Pauli-Villars condition
\[
\sum_{s\,=\,1}^S C_s =-1
\]
and consequently
\begin{eqnarray}
f(p^{\,2}) &=& -\,\frac{g^{\,2}}{16\pi^{2}}\int_{0}^{1}\mathrm{d}x\,x  
\sum_{n=1}^{\infty}\frac{1}{n!}(1-x)^{n}M^{2n}\sum_{s\,=\,1}^S C_s\lambda_{s}^{2n}
\int_{0}^{\infty}\frac{\mathrm{d}\tau}{\tau}\,\tau^{\,n}
\exp \left\lbrace \tau x(1-x)\,p^{\,2}\right\rbrace\0
\\
&=&\frac{g^{\,2}}{16\pi^{2}}\sum_{s\,=\,1}^S C_s\int_{0}^{1}\mathrm{d}x\,x \,
\ln\left( 1+\frac{\lambda_{s}^{2}M^{2}}{-\,xp^{\,2}}\right)\0
\\
&=&\frac{g^{\,2}}{16\pi^{2}}\left\lbrace 
-\,\frac12\,\ln\left( -\,\frac{p^{\,2}}{M^{2}}\right) - \int_{0}^{1}\mathrm{d}x\,x\,\ln x
+ \frac12\sum_{s\,=\,1}^S C_s\,\ln\lambda_{s}^{2}\right. \0
\\
&&\quad\quad\quad+ \left. \int_{0}^{1}\mathrm{d}x\,x\,\sum_{s\,=\,1}^S C_s\,\ln\left( 1 -\,\frac{xp^{\,2}}{M_{s}^{2}}\right)  
\right\rbrace \0
\\
&=&\left( \frac{g}{4\pi}\right)^{2} \left[\,
\sum_{s\,=\,1}^S C_s\,\ln\lambda_{s} + \frac14 + \frac12\ln\left( -\,\frac{M^{\,2}}{p^{\,2}}  \right)\, 
\right] + \mathrm{evanescent}\label{fp2PV}
\end{eqnarray}
It is quite instructive and enlightening a comparison
of this result with the corresponding ones derived from differente regulators.

\medskip
\textsc{dimensional and ultraviolet cut-off regularizations}

\medskip\noindent
In a $ 2\omega- $dimensional space-time the radiative correction to the Weyl kinetic term  takes the form 
\[
\mathtt{reg}\,\Sigma_2(p\!\!/) 
= -\,ig^{\,2}\mu^{\,2\epsilon}\int\frac{\mathrm{d}^{2\omega}\ell}{(2\pi)^{2\omega}}\;D_{\mu\nu}(\ell)\,
\gamma^{\,\mu}\,S(\ell+p)\,\gamma^{\,\nu}P_L
\]
where $ \epsilon=2-\omega>0 $ is the shift with respect to the Minkowski space. Since the above expression  
is traceless and has the canonical engineering dimension of a mass in natural units, it is quite apparent that the
latter cannot generate any mass term, which would be proportional to the unit matrix. 
Hence, mass is forbidden and it is convenient to set and evaluate
\begin{eqnarray*}
\mathtt{reg}\,\Sigma_2(p\!\!/) &\equiv& f(\,p^{\,2})\,p\!\!/P_L\qquad\quad
\mathrm{tr}\,[\,p\!\!/\mathtt{reg}\,\Sigma_2(p\!\!/)]=\textstyle\frac12\,2^{\,\omega}p^{\,2}f(\,p^{\,2})
\\
\mathrm{tr}\,[\,p\!\!/\mathtt{reg}\,\Sigma_2(p\!\!/)]
&=& g^{\,2}\mu^{\,2\epsilon}\,{(2\pi)^{-\,2\omega}}
\int{\mathrm{d}^{2\omega}\ell}\;
\frac{(-\,i\,)\,\mathrm{tr}\left(\,p\!\!/\gamma^{\,\lambda}\ell\!\!/\gamma_{\lambda}P_L\,\right) }
{\left[\,(\ell-p)^2+i\varepsilon\,\right]\left(\,\ell^{\,2}+i\varepsilon\,\right)}
\end{eqnarray*}
Explicit calculation and expansion around the shift off the Minkoski space $ 4-2\omega=2\epsilon\rightarrow0 $ yields
\begin{eqnarray}
f(p^2) &=&\left( \frac{g}{4\pi}\right)^{2} \left[\,\frac{1}{\epsilon} + 1
-\,\mathbf{C} + \ln\left( -\,\frac{4\pi\mu^{2}}{p^{\,2}}\right) \,\right] + \mathrm{evanescent}\label{fp2dim}
\end{eqnarray}

It follows that by identifying 
\[
\frac{1}{\epsilon}=\sum_{s\,=\,1}^S C_s\,\ln\lambda_{s} 
\]
in which the two Pauli-Villars conditions\footnote{This means the the minimal choice is $ S=2 $.} hold true, viz.,
\[
\qquad\sum_{s\,=\,1}^S C_s =-1\qquad\sum_{s\,=\,1}^S C_s\lambda_{s}=0
\]
the coefficient of the divergent part turns out to be the same for both PV and DR regulators, as expected.

\medskip\noindent
Moreover, it is quite instructive to repeat once again the 1-oop calculation
of the Weyl spinor self-energy with a physical very large cut-off regulator, e.g.  $ K=l_{P}^{-1} $ with $ l_{P}=\sqrt{\hslash G_N/c^{\,3}} $
the Planck length. Of course, once again, no mass term is allowed for the trace of an odd number of gamma matrices is null.
Moreover, for the previously introduced coefficient of the Weyl kinetic term we find with
$ \ell^{\,\mu}=(\ell_{0},	\vec{\ell}\;) $
\begin{eqnarray*}
p^2\,f(\,p^{\,2})  &=& \frac{-\,ig^{\,2}}{(2\pi)^{4}}
\int\frac{2p\cdot\ell\ \theta\left(K^{2}-\vec{\ell}^{\;2}\,\right)\,\mathrm{d}^{4}\ell}
{\left[\,(\ell - p)^2+i\varepsilon\,\right]\left(\,\ell^{\,2}+i\varepsilon\,\right)}
\\
&=& \frac{-\,ig^{\,2}}{(2\pi)^{4}}\int_{0}^{1}\mathrm{d}x
\int\frac{2p\cdot\ell\ \theta\left(K^{2}-\vec{\ell}^{\;2}\,\right)\,\mathrm{d}^{4}\ell}
{\left[\,\ell^{\,2} -2x p\cdot\ell + x p^{\,2}+i\varepsilon\,\right]^{2}}
\\
&=& \frac{-\,ig^{\,2}}{(2\pi)^{4}}\int_{0}^{1}\mathrm{d}x\int\mathrm{d}\vec{\ell}\ \theta\left(K^{2}-\vec{\ell}^{\;2}\,\right)
\lim_{\eta\,\rightarrow\,0}\,\frac{\mathrm{d}}{\mathrm{d}\eta}\int_{-\infty}^{\infty}\mathrm{d}\ell_{0}\;
\frac{ 2p_{0}\ell_{0}-2\vec{p}\cdot\vec{\ell} }
{\ell^{\,2} - 2x p\cdot\ell + x p^{\,2} -\eta +i\varepsilon}
\end{eqnarray*}
For a very large cut-off $ K\rightarrow\infty $ elementary calculations yield
\begin{eqnarray}
f(\,p^{\,2}) &=&\frac{g^{\,2}}{4\pi^{2}}\int_{0}^{1}\mathrm{d}x\,x\left\lbrace 
-1+\ln2+\frac12\ln\left( -\,\frac{4K^{2}}{p^{\,2}}\right) -\frac12\ln x(1-x)\right\rbrace +\cdots\0
\\
&=& \left( \frac{g}{4\pi}\right) ^{2}\left[ \,\ln\left( -\,\frac{4K^{2}}{p^{\,2}}\right) + \ln4 \,\right] 
+\mathrm{evanescent}\label{fp2cutoff}
\end{eqnarray}

\medskip
To sum up, we have verified that the 1-loop correction to the (left) Weyl spinor self-energy has the general form,
which is \textbf{universal, i.e. regularization independent}: namely,
\begin{eqnarray*}
\mathtt{reg}\,\Sigma_2(p\!\!/) &\equiv& f(\,p^{\,2})\,p\!\!/P_L
\\
&:=&\left( \frac{g}{4\pi}\right)^{2} \left[\,
\sum_{s\,=\,1}^S C_s\,\ln\lambda_{s} + \frac14 + \frac12\ln\left( -\,\frac{M^{\,2}}{p^{\,2}}  \right)\, 
\right]\qquad (\,\mathrm{PV}\,)
\\
f(\,p^{\,2})&:=&\left( \frac{g}{4\pi}\right)^{2} \left[\,\frac{1}{\epsilon} + 1
-\,\mathbf{C}+ \ln\left( -\,\frac{4\pi\mu^{2}}{p^{\,2}}\right) \,\right]\qquad  (\,\mathrm{DR}\,)
\\
&:=&\left( \frac{g}{4\pi}\right)^{2} \left[\,\ln\left( -\,\frac{4K^{2}}{p^{\,2}}\right) + \ln4 \,\right] 
\qquad\quad(\,\mathrm{UV}\,)
\end{eqnarray*}

\medskip\noindent
A few final remarks and comments are in  order.
\begin{enumerate}
\item In the present model of a left-handed Weyl spinor minimally coupled to a gauge vector potential,
\textbf{no mass term can be generated by the radiative corrections} in any regularization scheme.
The left-handed part of the classical kinetic term does renormalize, while its right-handed part
does not undergo any radiative correction and remains free. The latter has to be necessarily 
introduced in order to define a Feynman propagator for the massless spinor field, like the gauge
fixing term to invert the kinetic term of the gauge potential (see \eqref{bardeen}).
The (one loop) renormalized  Lagrangian for a Weyl fermion minimally coupled with a gauge vector potential 
has the universal - i.e. regularization independent - form
\begin{eqnarray*}
\mathcal{L}_{1-\rm loop} &=& \mathcal{L}=\chi_{L}^{\dagger}\,\alpha^{\nu}i\partial_{\nu}\chi_{L}
+ gA^{\nu}\,\chi_{L}^{\dagger}\,\alpha_{\nu}\chi_{L} -\textstyle\frac14\,F^{\,\mu\nu}\,F_{\mu\nu}
\\
&+& \varphi_{R}^{\dagger}\,\alpha^{\nu}i\partial_{\nu}\varphi_{R}  
-\textstyle\frac12(\partial\cdot A)^{2} - (Z_{3}-1)\textstyle\frac14\,F^{\,\mu\nu}\,F_{\mu\nu}
\\
&+& (Z_{2}-1)\chi_{L}^{\dagger}\,\alpha^{\nu}i\partial_{\nu}\chi_{L}
+ (Z_{1}-1)gA^{\nu}\,\chi_{L}^{\dagger}\,\alpha_{\nu}\chi_{L} 
\\
(Z_{2}-1)&=& -\left( \frac{g}{4\pi}\right)^{2} \left[\,\frac{1}{\epsilon} + F_{2}(\epsilon,p^{\,2}/\mu^{2})\,\right] 
+ \cdots
\\
&=& -\left( \frac{g}{4\pi}\right)^{2} \left[\,
\sum_{s\,=\,1}^S C_s\,\ln\lambda_{s} + \widetilde F_2(\lambda_{s},p^{\,2}/M^{2})\,\right] 
+ \cdots
\\
&=& -\,\left( \frac{g}{4\pi}\right)^{2} \left[\,\ln\left( -\,\frac{4K^{2}}{p^{\,2}}\right) + \ln4 + \widehat{F}_{2}(K^{2}/p^{\,2})\,\right] 
+\cdots
\end{eqnarray*}
where conventional notations have been employed. Notice that, as usual, the arbitrary finite parts $ F_2,\widetilde F_2, \widehat{F}_2 $ 
of the countertems  are analytic for $ \epsilon\rightarrow0 $ and $ \lambda_{s},K\rightarrow\infty $, respectively,  and have to be 
uniquely fixed by the renormalization prescription.
\item The interaction definitely preserves left chirality and scale invariance of the counterterms  in the transition from the classical to the 
(perturbative) quantum theory: no mass coupling between the left-handed (interacting) Weyl spinor $ \chi_{L} $
and right-handed (free) Weyl spinor $ \varphi_{R} $ can be generated by radiative loop corrections.

\item While the cut-off and dimensional regularized theory does admit a local formulation 
in  $ D=4 $ or $ D=2\omega $ space-time dimensions, there
is no such local formulation for the Pauli-Villars regularization. The reason is that the PV spinor propagator
\[
\sum_{s\,=\,0}^S C_s\,S(\ell\,,\,M_s)
\]
where $M_0 = 0\,,\ C_0=1$ while 
$\{\,M_s\equiv \lambda_s\,M\,|\,\lambda_s\gg 1\ (\, s=1,2,\ldots,S\,)\,\}$,
cannot be the inverse of any local differential operator of the Calderon-Zygmund type.
Hence, there is no local action involving a bilinear spinor term that can produce, after a suitable causal inversion,
the Pauli-Villars regularized spinor propagator.
\end{enumerate}

\medskip\noindent
It follows therefrom that the one loop effective Lagrangian proposed in \cite{Bast1,Bast2} has nothing to do with the Pauli-Villars regularization of the present model.

\medskip
Now the main point concerning the chiral anomaly and unitarity. 
As we have seen in detail, in the present model of a left-handed Weyl spinor
interacting with a gauge vector potential, all the symmetries of the classical action do survive in the divergent and finite parts of the
counter-terms, for any choice of the regulators and renormalization prescription. As it is well-known since long, this is not 
so for the famous chiral left-handed triangle diagram, which turns out to be UV finite and yields
\begin{eqnarray*}
\left.\tilde\Gamma^{(L)}_{\mu_1\mu_2\mu_3}(k_1,k_2,k_3)\right|_{\rm 1-loop}
&=& \frac{g^{\,3}}{12\pi^2}\,\varepsilon_{\sigma\mu_1\mu_2\mu_3}	\,
\left(k_1^\sigma -k_2^\sigma\right)\,k_3^2\, I_{12}(k_1,k_2,k_3) +
{\rm cyclic\ permutations}
\\
&+& \frac{g^{\,3}}{4\pi^{2}}\,\varepsilon_{\sigma\tau\mu_1\mu_2}\,k_1^\sigma k_2^\tau k_3^{\mu_3}\,
I_{12}(k_1,k_2,k_3)+{\rm cyclic\ permutations}
\\
&&(\,k_1+k_2+k_3=0\,)
\\
I_{rs}(k_1,k_2,k_3)
&=& -\int_0^1 \mathrm{d} x_1\int_0^1	\mathrm{d} x_2\int_0^1\mathrm{d} x_3\,
\frac{2x_r\,x_s\,\delta(1-x_1-x_2-x_3)}{k_1^2 x_2 x_3+k_2^2 x_3 x_1+k_3^2 x_1 x_2}
\\
&&(\,r,s=1,2,3\,)
\end{eqnarray*}
The one loop chiral left-handed triangular amplitude obeys the anomalous Ward identity
\[
k_1^\sigma\,\tilde\Gamma^{(L)}_{\sigma\mu_2\mu_3}(k_1,k_2,k_3)=
-\,\frac{g^{\,3}}{12\pi^{2}}\,\varepsilon_{\sigma\tau\mu_2\mu_3}\,k_2^\sigma\,k_3^\tau
\]
and related cyclic permutations. 
\\
\noindent
The above expression for the 1-loop triangular amplitude is finite, it does not
depend at all upon regulators, or counterterms or even renormalization prescriptions. However, it turns out that, owing to the anomalous Ward identity, the left-handed Weyl current
does no longer satisfy the continuity equation, at variance with the classical case, viz.,
\begin{equation}
\partial_{\nu}\chi_{L}^{\dagger}\,\alpha^{\,\nu}\chi_{L} = \frac13\left( \frac{g}{4\pi}\right) ^{2}F^{\,\mu\nu}_{\ast}F_{\mu\nu}
\qquad\qquad F^{\,\mu\nu}_{\ast}=\textstyle\frac12\,\varepsilon^{\,\mu\nu\rho\sigma}\,F_{\rho\sigma}\label{Abelianconsistanom}
\end{equation}
This anomalous equation is written in the covariant form. But this anomaly is what, in the non Abelian case, is known as consistent gauge anomaly. This is due to the fact that, as already pointed out, in the Abelian case covariant and consistent gauge anomalies have the same form. However the coefficient of the anomaly in \eqref{Abelianconsistanom} is characteristic of consistent anomalies in 4d, see \eqref{A}. The anomaly in  \eqref{Abelianconsistanom} is in fact the split consistent U(1) anomaly (changing chirality changes its sign), and, like all the consistent gauge anomalies, it puts in jeopardy unitarity. As a matter of fact the Lorentz invariant quantum theory of a gauge vector field unavoidably
involves a Fock space of states with indefinite norm. Now, in order to select a physical Hilbert subspace of the
Fock space a subsidiary condition is necessary. In the Abelian case, when the fermion current satisfies the continuity equation then 
the equations of motion lead to  $ \square(\partial\cdot A)=0 $, so that a subspace of states of non-negative norm can be selected
through the auxiliary condition
\[
\partial\cdot A^{(-)}(x)\,\vert\,\mathrm{phys}\,\rangle=0
\]
$A^{(-)}(x)$ being the destruction, positive frequency part of a d'Alembert quantum field.
Conversely, in the present chiral model we find
\[
-\,g^{-1}\square(\partial\cdot A)=\partial_{\nu}\chi_{L}^{\dagger}\,\alpha^{\,\nu}\chi_{L} 
= \frac13\left( \frac{g}{4\pi}\right) ^{2}F^{\,\mu\nu}_{\ast}F^{\mu\nu}\neq0
\]
in such a manner that nobody knows how to select a physical subspace of states with non negative norm, if any, where a unitary restriction of the collision operator $ S $ could be defined.

\vskip 1cm


\end{document}